\documentclass[prc,aps,floatfix,nofootinbib,twocolumn]{revtex4} 

\usepackage{graphicx}
\usepackage{color}
\usepackage{enumerate}
\usepackage{amssymb}
\usepackage{amsmath}
\usepackage{amsfonts}
\usepackage{bbm}
\usepackage{dsfont}
\def\be{\begin{equation}} 
\def\ee{\end{equation}}

\begin{document}

\title{
A simple schematic model for a cross section 
deficit in $^{12}$C+$^{12}$C fusion reactions }

\author{K. Hagino}
\affiliation{ 
Department of Physics, Kyoto University, Kyoto 606-8502,  Japan} 

\begin{abstract}
A cross section deficit phenomenon has been observed 
in $^{12}$C+$^{12}$C fusion reactions at astrophysical energies, 
at which fusion cross sections are suppressed in the off-resonance regions as compared to fusion cross 
sections for the $^{12}$C+$^{13}$C system. 
I here construct a simple schematic model which simulates 
this phenomenon. The model consists of a random matrix Hamiltonian based on the Gaussian 
Orthogonal Ensemble (GOE), which is coupled to an entrance 
channel Hamiltonian in the discrete basis representation. 
I show that the transmission coefficients are almost unity when both the level density and the 
decay widths of the GOE 
configurations are large, realizing the strong absorption regime. 
On the other hand, when these parameters are small, the transmission coefficients are significantly structured 
as a function of energy. In that situation, the transmission coefficients at resonance energies 
reach unity in this model, that is consistent with the experimental finding of the cross section deficit. 
\end{abstract}

\maketitle

\section{Introduction} 

The $^{12}$C+$^{12}$C fusion reaction at extremely low energies has attracted lots of 
attention both experimentally and theoretically \cite{patterson1969,kovar1979,satkowiak1982,high1977,
notani2012,zhang2020,spillane2007,tan2020,fruet2020,jiang2013,tumino2018,imanishi1968,
imanishi1969,kondo1978,scheid1970,fink1972,esbensen2011,assuncao2013,diaz-torres2018,bonasera2020,
taniguchi2021}, largely because this is one of the key reactions 
in nuclear astrophysics. This reaction is important e.g., in carbon burning in massive stars, type Ia supernovae, and X-ray superburst. A characteristic feature of the fusion cross sections of this system is that there are 
many prominent resonance peaks. This is in contrast to fusion cross sections of a 
similar system, the $^{12}$C+$^{13}$C fusion reaction, 
which show a much smoother energy dependence of cross sections. Another interesting observation is that the fusion cross sections 
of the $^{12}$C+$^{12}$C reaction at resonance energies appear to coincide with the cross sections of the 
$^{12}$C+$^{13}$C reaction, while those at off-resonance energies are relatively suppressed  \cite{notani2012,zhang2020}. 
That is, the fusion cross sections of the $^{12}$C+$^{13}$C reaction
seem to provide the upper limit of the fusion cross sections of 
the $^{12}$C+$^{12}$C system. 
This was referred to as a cross section 
deficit in Ref. \cite{jiang2013}. Its mechanism has not yet been completely clarified. 

It was argued in Ref. \cite{jiang2013} that the different behaviors of fusion 
cross sections in these systems are caused by i) a lower $Q$-value of the $^{12}$C+$^{12}$C fusion reaction 
as compared to the $Q$ value of the $^{12}$C+$^{13}$C fusion reaction, 
ii) the compound nucleus formed in the $^{12}$C+$^{12}$C fusion reaction is $^{24}$Mg, that is an even-even 
nucleus and thus the level density is lower than that of $^{25}$Mg formed in the $^{12}$C+$^{13}$C reaction, 
and iii) $^{12}$C+$^{12}$C is a reaction of identical bosons so that only states with positive parity and 
even spin in $^{24}$Mg can be populated. All of these lead to the fact that 
resonances are isolated in the $^{12}$C+$^{12}$C fusion reaction at low energies, while the $^{12}$C+$^{13}$C fusion reaction is in the overlapping resonance regime 
even at low energies. In the former case, it is known that average transmission coefficients are suppressed 
by the Moldauer factor, $P_J=1-e^{-2\pi\Gamma_J/D_j}$, where $J$ is the spin of the compound nucleus, $\Gamma_J$ 
is the average decay width, and $D_J$ is the average level spacing \cite{jiang2013,moldauer1967,moldauer1968,celardo2008}. 
See also Refs. \cite{moldauer1969,simonius1974} for the derivation of the Moldauer factor. 
Notice that the low $Q$-value of the the $^{12}$C+$^{12}$C fusion reaction
also leads to the fact that the neutron emission 
channel from the compound nucleus is closed in the $^{12}$C+$^{12}$C fusion reaction, 
while such channel is open 
in the $^{12}$C+$^{13}$C fusion reaction. 

The Moldauer factor explains why the fusion cross sections of the $^{12}$C+$^{12}$C system are 
suppressed on average as compared to the cross sections of the $^{12}$C+$^{13}$C system. However, it does not 
provide a reasoning why the fusion cross sections of the $^{12}$C+$^{12}$C system at the resonance energies 
coincide with the cross sections of the $^{12}$C+$^{13}$C system. 
To address this question, in this paper I consider a simple schematic model which was used in Appendix B 
of Ref. \cite{bertsch2023}. The model consists of a random matrix Hamiltonian 
based on the Gaussian Orthogonal Ensemble (GOE) which is coupled to scattering states in the entrance  
channel. Each of the GOE configurations posses a decay width $\gamma$ 
so that a part of the incident flux is absorbed 
by the GOE configurations. By changing the parameters in the GOE Hamiltonian and the decay width $\gamma$, one 
can simulate both the isolated resonance regime and the overlapping resonance regime, and thus 
the model provides a convenient tool to qualitatively understand the difference in the fusion cross sections 
of the two C+C systems. 

The paper is organized as follows. In Sec. II, I detail the simple schematic model which is employed in this 
paper. In particular, I explain in detail how one can compute the transmission coefficients, or the absorption 
probabilities, based on the discrete basis formalism. In Sec. III, I compare the isolated resonance regime 
and the overlapping resonance regime with this model and show that the model can qualitatively reproduce the experimental findings in 
the C+C fusion reactions. I then summarize the paper in Sec. IV. In the Appendix, I detail a numerical 
method to compute 
poles of the $S$-matrix in the complex energy plane. 

\section{Model}

\begin{figure}[htb] 
\begin{center} 
\includegraphics[width=0.9\columnwidth]{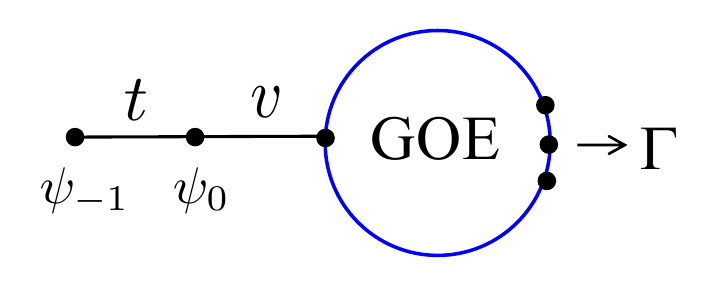} 
\caption{
A schematic illustration of the model Hamiltonian 
employed in this paper. It consists of a random matrix 
based on the Gaussian Orthogonal Ensemble (GOE), which 
is coupled to the free scattering states ${\psi_i}$ 
with the strength $v$. 
$\Gamma$ is the decay width of the GOE configurations, while 
$t$ denotes the off-diagonal 
component of the kinetic energy operator, which connects 
neighboring scattering states in the discrete basis 
representation. 
}
\label{fig:model}
\end{center} 
\end{figure} 

The model employed in this paper 
is schematically illustrated in Fig. 1. 
This is the same model 
as that in Appendix B in Ref. \cite{bertsch2023}, 
and consists of a GOE Hamiltonian $H_{\rm GOE}$ 
with the dimension 
of $N_{\rm GOE}$. 
The matrix elements of the GOE Hamiltonian read,
\begin{equation}
    (H_{\rm GOE})_{ij}=v_g r_{ij} \sqrt{1+\delta_{i,j}},
\end{equation}
where $v_g$ is the average strength of the 
interaction and $r_{ij}$ is a random number from a Gaussian distribution
of unit dispersion, $\langle r_{ij}^2\rangle =1$. 
I assume that the first $N_{\rm decay}$ basis 
states have a decay width of $\gamma$. 
Thus the total GOE Hamiltonian is given by 
\begin{equation}
H_{\rm GOE}'=H_{\rm GOE}-\frac{i}{2}\Gamma,
\end{equation}
where $\Gamma$ is a diagonal matrix in which the 
first $N_{\rm decay}$ components are $\gamma$ and the rest is zero. 

In Fig. 1, $\psi_0$ and $\psi_{-1}$ represent 
free scattering wave functions at $x_0$ and $x_{-1}~(=x_0-\Delta x)$, respectively, $\Delta x$ being 
the mesh spacing. 
The free scattering wave functions satisfy 
the equation 
\begin{equation}
t\psi_{i-1}+t\psi_{i+1}=(E'+2t)\psi_i\equiv E\psi_i, 
\end{equation}
where $\psi_i$ is the wave function at $x_i$. Here the scattering energy 
$E'$ is shifted as $E=E'+2t$ so that the diagonal component of the kinetic energy 
operator, $-2t$, does not appear in the equation. 
Assuming the 
plane wave solution, $\psi_i\propto e^{\pm ikx_i}$, 
one finds that the relation between $E$ and $k$ 
is given by 
\begin{equation}
    E=2t\cos(k\Delta x). 
    \label{dispersion}
\end{equation}

I assume that the free wave function $\psi_0$ 
couples to the first basis state of the GOE Hamiltonian 
with the strength $v$. 
Thus, the wave functions $\psi_0$ and $\psi_{-1}$, 
as well as the amplitudes $\{f_i\}$ 
for the GOE basis states satisfy the equation
\begin{equation}
\left(\begin{matrix}
0 & \vec{v}^T \\
\vec{v} & H_{\rm GOE}'
\end{matrix}
\right)
\left(\begin{matrix}
\psi_0 \\
\vec{f}
\end{matrix}
\right)
=
E
\left(\begin{matrix}
\psi_0 \\
\vec{f}
\end{matrix}
\right)
+
\left(\begin{matrix}
-t\psi_{-1} \\
\vec{0}
\end{matrix}
\right), 
\label{eq:Hpsi}
\end{equation}
where the $N_{\rm GOE}$-component vectors $\vec{0}$ and $\vec{v}$ are 
defined as $\vec{0}=(0,0,\cdots,0)^T$ and 
$\vec{v}=(v,0,\cdots,0)^T$, respectively. 
Here, $T$ denotes the transpose. 
By taking the last $N_{\rm GOE}$ rows in Eq. (\ref{eq:Hpsi}), 
\begin{equation}
(H_{\rm GOE}'-E)\vec{f}=
\left(\begin{matrix}
-v\psi_0 \\
0 \\
\vdots \\
0
\end{matrix}
\right), 
\end{equation}
the amplitudes $\vec{f}$ can 
be obtained as a ratio to $\psi_0$ as \cite{fanto2018,hagino2024}
\begin{equation}
    f_i=-(H'_{\rm GOE}-E)_{i1}^{-1}v\psi_0
    \equiv -G_{i1}v\psi_0, 
\end{equation}
where $G\equiv (H'_{\rm GOE}-E)^{-1}$ is the Green function. 
After the amplitudes $\vec{f}$ are so obtained, the 
wave function $\psi_{-1}$ is obtained as 
\begin{equation}
\psi_{-1}=-\frac{1}{t}(-E\psi_0+vf_1)
=-\frac{\psi_0}{t}(-E-G_{11}v^2). 
\end{equation}
Since $\psi_0$ and $\psi_{-1}$ represent the free scattering 
wave functions, they should satisfy 
the relation
\begin{eqnarray}
\psi_{-1}&=&A(e^{-ik\Delta x} -S e^{ik\Delta x}), \\
\psi_0&=&A(1 -S), 
\end{eqnarray}
where $S$ is the $S$-matrix. From these equations, the $S$-matrix reads, 
\begin{equation}
    S=\frac{\psi_{-1}-e^{-ik\Delta x}\psi_0}
    {\psi_{-1}-e^{ik\Delta x}\psi_0}
    =\frac{E+G_{11}v^2-te^{-ik\Delta x}}
    {E+G_{11}v^2-te^{ik\Delta x}}. 
    \label{S-matrix}
\end{equation}
From the $S$-matrix, the absorption probability, or 
the transmission coefficient, is obtained as
\begin{equation}
    T=1-|S|^2. 
\end{equation}

\section{Role of GOE properties in transmission 
coefficients}

\begin{figure}[tb] 
\begin{center} 
\includegraphics[width=0.9\columnwidth]{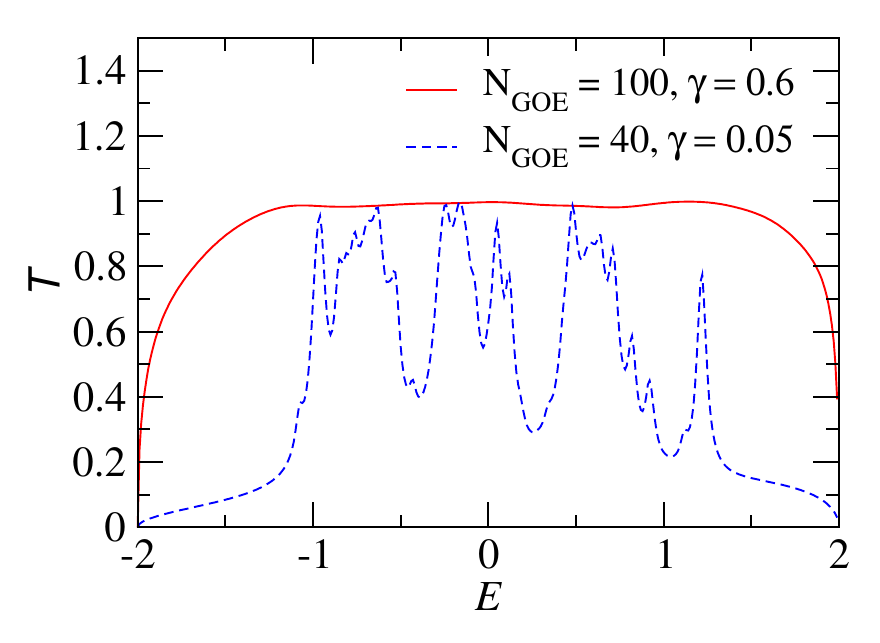} 
\caption{
Transmission coefficients obtained 
with a single random seed. 
The red solid and the blue dashed lines are obtained 
with $(N_{\rm GOE},\gamma)=(100,0.6)$ and 
$(N_{\rm GOE},\gamma)=(40,0.05)$, respectively. 
The number of decay channel, $N_{\rm decay}$, is
set to be $N_{\rm decay}=N_{\rm GOE}$ for each case. 
The other parameters are 
$v_g=0.1$, $t=-1$, and 
$v=1$. 
}
\label{fig:t}
\end{center} 
\end{figure} 

Let us now numerically evaluate the transmission coefficients 
and discuss the role played by the parameters in the 
GOE Hamiltonian. The red solid and the blue dashed lines 
in Fig. \ref{fig:t} show the transmission coefficients obtained 
with a single random seed 
for the parameters $(N_{\rm GOE},\gamma)=(100,0.6)$ and 
$(N_{\rm GOE},\gamma)=(40,0.05)$, respectively. To this end, I set $N_{\rm decay}=N_{\rm GOE}$ and take $v_g=0.1$, $t=-1$, and 
$v=1$. 
Notice that the 
transmission coefficients drop to zero at $E=\pm 2$, 
because the absolute value of energy $E$ cannot be larger than 2$t$ due to the dispersion relation, Eq. (\ref{dispersion}). Nevertheless, one can still discuss 
the behavior of the transmission coefficients around 
the central energy, $E\sim 0$. 
In the case of $(N_{\rm GOE},\gamma)=(100,0.6)$, the 
transmission coefficients are close to unity. This is 
consistent with the strong absorption limit, which 
is often assumed in heavy-ion 
reactions \cite{hagino2022,hagino2012}. 
On the other hand, in the case of 
$(N_{\rm GOE},\gamma)=(40,0.05)$, the transmission 
coefficients are much more structured as a function 
of $E$ with several prominent resonance peaks. It is interesting to observe that the transmission 
coefficients are close to unity at a few resonance 
energies. This is similar to the cross section deficit 
observed in the $^{12}$C+$^{12}$C reaction. The present simple model thus captures an essential feature of the C+C fusion reactions. 

\begin{figure}[t] 
\begin{center} 
\includegraphics[width=0.9\columnwidth]{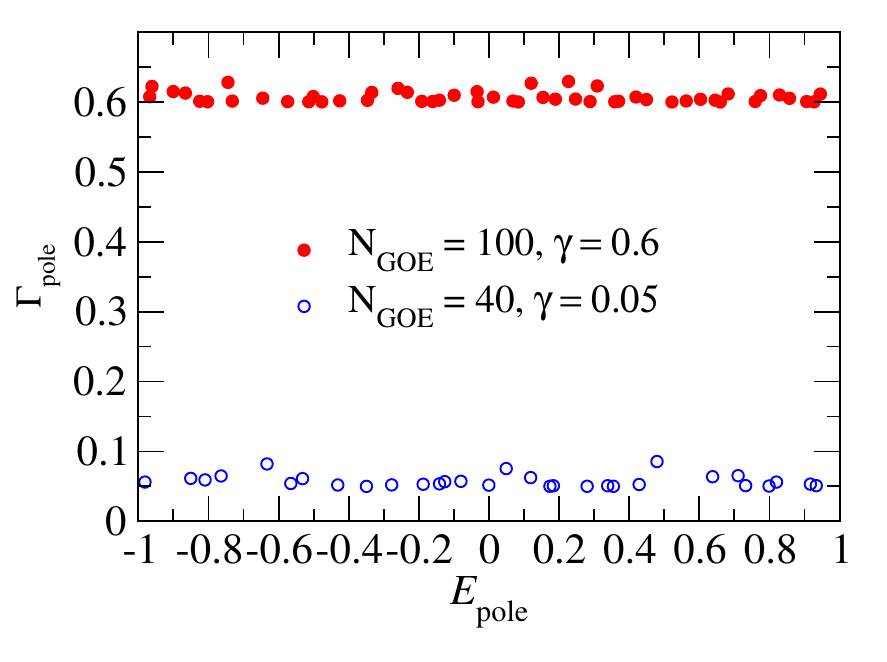} 
\caption{
The energies of the resonance poles, 
$E_{\rm pole}-i\Gamma_{\rm pole}/2$, 
corresponding to the two curves in Fig. \ref{fig:t}. 
}
\label{fig:pole}
\end{center} 
\end{figure} 

The complex energies, 
$E_{\rm pole}-i\Gamma_{\rm pole}/2$, for the 
corresponding $S$-matrix poles are shown in Fig. \ref{fig:pole}. See Appendix for the numerical 
method to evaluate the resonance poles. 
The widths $\Gamma_{\rm pole}$
fluctuate around the unperturbed width, $\gamma$. 
The average level spacing is also close to the 
inverse of the level density of the GOE Hamiltonian, 
$\rho_{\rm GOE}=\sqrt{N_{\rm GOE}}/\pi v_g$. 
For $v_g=0.1$, 
this is $1/\rho_{\rm GOE}=0.03$ and 0.05 for 
$N_{\rm GOE}=100$ and 40, respectively. 
The value of $\gamma$ is much larger than 
$1/\rho_{\rm GOE}$ in the case of 
$(N_{\rm GOE},\gamma)=(100,0.6)$, 
and the resonances are largely overlapping. As a consequence, 
the transmission coefficients show a smooth energy dependence as shown in Fig. \ref{fig:t}. 
On the other hand, in the case of 
$(N_{\rm GOE},\gamma)=(40,0.05)$, the value of 
$\gamma$ is comparable to $1/\rho_{\rm GOE}$, 
leading to the structured transmission coefficients. 

\begin{figure}[t] 
\begin{center} 
\includegraphics[width=0.9\columnwidth]{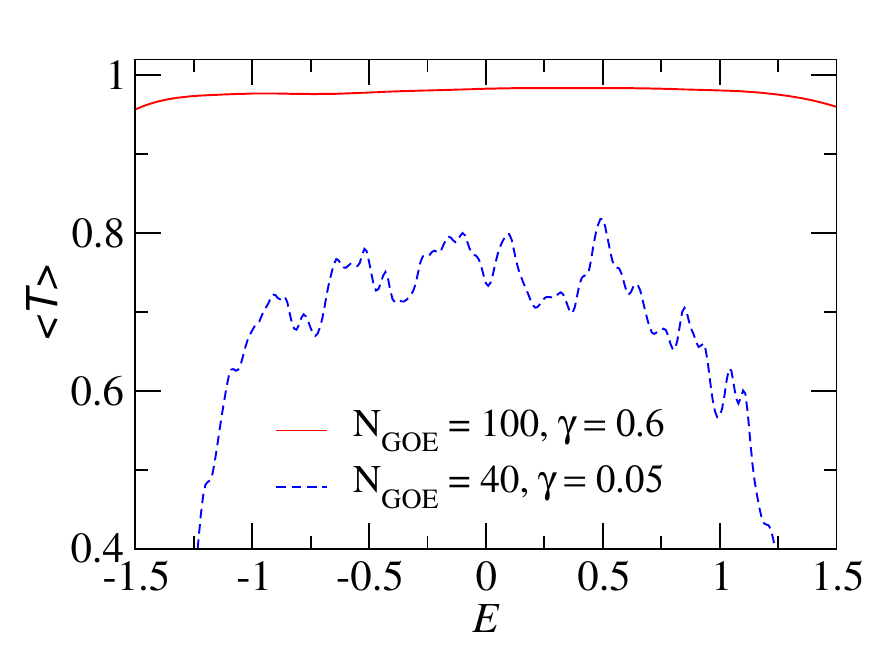} 
\caption{An ensemble average of the transmission 
coefficients, 
$\langle T(E)\rangle=1-\langle |S(E)|^2\rangle$, 
obtained with 20 ensembles. The 
parameters of the Hamiltonian are the same as 
those in Fig. \ref{fig:t}. 
}
\label{fig:t-av}
\end{center} 
\end{figure} 

An ensemble average of the transmission coefficients, 
\begin{equation}
    \langle T(E) \rangle = 1-\langle |S(E)|^2\rangle, 
    \label{eq:t-av}
\end{equation}
is shown in Fig. \ref{fig:t-av}. 
To this end, I take an average of 20 ensembles. 
Because of the ensemble average, the structure 
of the blue dashed curve is significantly smeared as 
compared to that shown in Fig. \ref{fig:t} for a single random seed. Yet, one can still observe that 
the dashed line is much more structured than 
the solid line. 

\begin{figure}[hbt] 
\begin{center} 
\includegraphics[width=0.9\columnwidth]{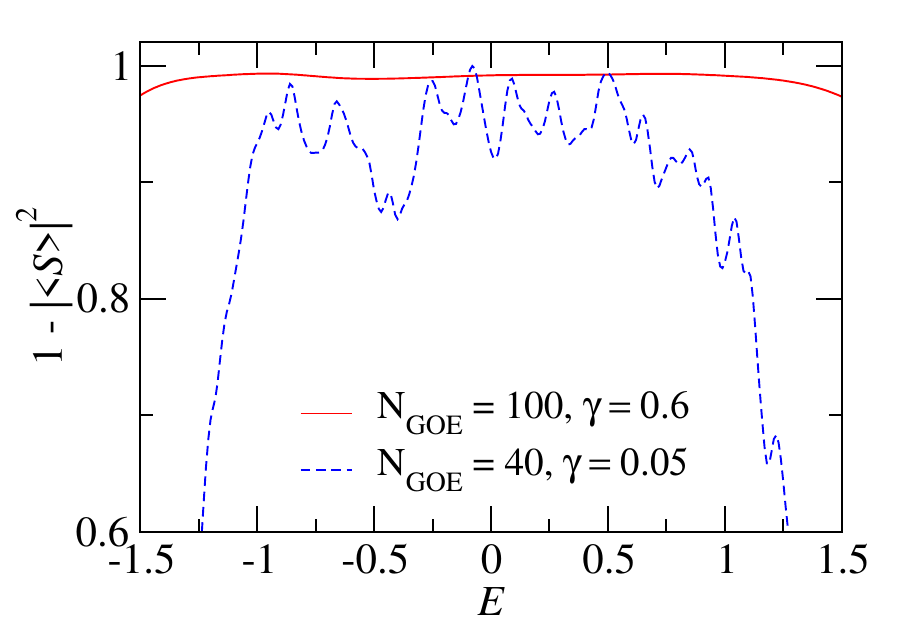} 
\caption{
The average transmission coefficients, 
$\bar{T}(E)=1-|\langle S(E)\rangle|^2$, 
obtained by firstly taking the ensemble average of the $S$-matrix. 
}
\label{fig:t-av2}
\end{center} 
\end{figure} 

In the compound nucleus theory, the transmission 
coefficients obtained by an optical potential 
correspond to 
\begin{equation}
    \bar{T}(E)=1-|\langle S(E)\rangle|^2.  
\end{equation}
That is, the ensemble average is taken 
firstly for the $S$-matrix and then take the absolute 
value square to obtain the average transmission coefficients. 
If one decomposes the average transmission coefficients (\ref{eq:t-av}) as,
\begin{eqnarray}
    \langle T \rangle 
    &=& 1-|\langle S(E)\rangle|^2 - 
    (\langle |S(E)|^2\rangle-|\langle S(E)\rangle|^2), \\ 
    &=&\bar{T}-(\langle |S(E)|^2\rangle
    -|\langle S(E)\rangle|^2), 
\end{eqnarray}
$\bar{T}$ corresponds to the probabilities of a compound nucleus formation, while 
the second term, $(\langle |S(E)|^2\rangle-|\langle S(E)\rangle|^2)$, corresponds to the compound-elastic 
scattering \cite{bertulani2004}. 
The average transmission coefficients $\bar{T}$ 
are plotted in Fig. \ref{fig:t-av2}. 
These transmission coefficients are qualitatively 
similar to the ensemble average of $T(E)$, even 
though the blue dashed line is somewhat increased. 
Interestingly, 
some of the resonance peaks are close to 
the solid line in the strong limit. 
Notice that the average transmission 
coefficients $\bar{T}$ are the quantities which can 
be directly compared to the Moldauer factor, 
$1-e^{-2\pi \Gamma\rho_{\rm GOE}}$. If one uses 
$\Gamma=0.05$ and $\rho_{\rm GOE}=1/0.05$ for $N_{\rm GOE}=40$, 
this factor is found to be 0.998. 
This deviates somewhat from the average transmission 
coefficients shown in Fig. \ref{fig:t-av2}, but 
this would be reasonable as the transmission coefficients would need to be small in order to validate 
the 
approximations used to derive the 
Moldauer factor \cite{moldauer1969,simonius1974}. 

\section{Summary}

I have constructed a simple schematic model 
for a compound nucleus reaction and discussed 
the dependence of the transmission coefficients 
on the properties of the compound nucleus. The model 
consists of a random matrix based on the Gaussian 
Orthogonal Ensemble (GOE), which couples to a free scattering wave. By changing the dimension of the 
GOE Hamiltonian and the decay width of the compound 
nucleus states, the model can describe both the 
overlapping resonance regime and the isolated 
resonance regime. 
I have demonstrated that the transmission 
coefficients in the isolated resonance regime 
are suppressed at off-resonance energies as 
compared to the transmission coefficients 
in the overlapping resonance regime, which are 
close to unity. At on-resonance energies, the 
transmission coefficients in the isolated resonance 
regime are close to unity, exhibiting the deficit 
phenomenon at off-resonance energies. This is qualitatively the same as the experimental observations 
in the fusion cross sections of 
the $^{12}$C+$^{12}$C and $^{12}$C+$^{13}$C systems.
Therefore, even though the present model is simple enough, it  
still captures an essential feature of the 
fusion reactions of two carbon nuclei. 

The present model can be extended by replacing 
the GOE Hamiltonian by a more 
realistic many-body Hamiltonian, such as a shell 
model Hamiltonian. Such study will shed light on 
the dynamics of resonance reactions relevant to 
nuclear astrophysics, while keeping a quantitative 
description of the cross sections. I will report on 
this in a separate publication. 

\begin{acknowledgments}
I thank G.F. Bertsch, X.D. Tang, A.B. Brown and K. Uzawa 
for useful discussions. 
This work was supported in part by
JSPS KAKENHI Grant Number JP23K03414.
\end{acknowledgments}

\appendix

\section{Poles of the $S$-matrix}

To find the poles of the $S$-matrix, Eq. (\ref{S-matrix}), 
in the complex plane, I closely follow the method in Supplemental 
Material of Ref. \cite{fanto2018}. In this method, 
the outgoing boundary condition 
$\psi_0=e^{-ik\Delta x}\psi_{-1}$ is imposed to the scattering wave functions, 
and an iterative procedure is introduced to find the 
poles. 
Substituting 
$\psi_{-1}=e^{ik\Delta x}\psi_{0}$ to Eq. (\ref{eq:Hpsi}), 
one finds 
\begin{equation}
M(x)\vec{\psi}\equiv (H+te^{ik\Delta x}C-E)\vec{\psi}=0, 
\label{Mpsi}
\end{equation}
where $x$ is defined as $x\equiv k\Delta x$, and 
$\vec{\psi}$ and $H$ are defined as 
$\vec{\psi}\equiv(\psi_0,\vec{f})^T$ and 
\begin{equation}
    H\equiv \left(\begin{matrix}
0 & \vec{v}^T \\
\vec{v} & H_{\rm GOE}'
\end{matrix}
\right), 
\end{equation}
respectively. $C$ is a matrix with the elements of 
$C_{ij}=\delta_{i,j}\delta_{i,0}$. 

If one takes an arbitrary number $x=x_g$, Eq. (\ref{Mpsi}) 
is not satisfied in general. 
However, one can improve $x_g$ by assuming that 
it is close to $x_r$ which satisfies Eq. (\ref{Mpsi}). 
From $M(x_r)\vec{\psi}=0$ and $x_r=x_g+x_r-x_g$, one 
obtains
\begin{equation}
    \left[M(x_g)+(x_r-x_g)\left.\frac{dM}{dx}\right|_{x_g}\right]\vec{\psi}=0, 
\end{equation}
up to the first order of $x_r-x_g$ 
This equation is transformed to 
\begin{equation}
    (M'(x_g))^{-1}M(x_g)\vec{\psi}=(x_g-x_r)\vec{\psi}. 
\end{equation}
Since the right-hand side of this equation vanishes for resonance poles, one can take the lowest modulus 
eigen-value $\bar{\lambda}$ of 
$(M'(x_g))^{-1}M(x_g)$ and update $x_g$ as 
\begin{equation}
x_{g+1}=x_g-\bar{\lambda}. 
\end{equation}
Notice that $M'(x)$ is given by 
$M'(x)=ite^{ix}C+2t\sin(x)$ with $E$ given by Eq. (\ref{dispersion}). 
This is a diagonal matrix and can easily be inverted. 

This iterative procedure has to start with a good initial 
guess of $x_g$. In this paper, I use the complex 
eigenvalues $E_i$ of the matrix 
$H'_{\rm GOE}=H_{\rm GOE}-i\Gamma/2$ and convert them 
to $x_g^{(i)}=\cos^{-1}(E_i/2t)$.

\end{document}